*https://doi.org/10.23913/ricea.v10i19.159*
*Artículos Científicos*# El tequila para consumo nacional como una ventana de oportunidades para el pequeño productor agavero

*Tequila for national consumption as a window of opportunity for the small agavero producer*

*Tequila para consumo nacional como janela de oportunidade para o pequeno produtor de agavero*

**Guillermo José Navarro del Toro**
Universidad de Guadalajara, Centro Universitario de los Altos, México
guillermo.ndeltoro@gmail.com
guillermo.ndeltoro@academicos.udg.mx
https://orcid.org/0000-0002-4316-879X
**Resumen**

El objetivo de la presente investigación fue determinar el grado de conocimiento que los pobladores de la Zona Metropolitana de Guadalajara (integrada por los municipios Guadalajara, Tlajomulco de Zúñiga, Tlaquepaque, Zapopan y Tonalá) tenían en cuanto al tequila y a las marcas producidas en los Altos de Jalisco. Para ello, se diseñó una encuesta conformada por cinco preguntas, la cual fue aplicada en la plaza central, centro o zócalo de cada municipio. Los resultados demuestran que la grandes marcas al ser adquiridas por compañías internacionales, enfocaron su atención a la captura del consumidor en los mercados internacionales, ya que los precios que tienen los mismos productos que exportan ha quedado fuera del alcance del bolsillo de aquellos que gustan de esa bebida, por lo que se podría considerar que las grandes marcas, han dejando un poco atrás el mercado nacional por supuesto no lo abandonaron del todo, pero dejó de ser su principal objetivo. Por tanto, se puede concluir que el mercado nacional es la ventana de oportunidad de conjuntar a los pequeños y aún desconocidos productores para trabajando conjuntamente y de forma
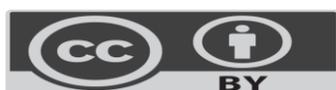

**Vol. 10, Núm. 19     Enero – Junio 2021**

agrupada, sean capaces de estandarizar una serie de productos que siendo de la misma calidad y mismos envases, puedan llegar a cubrir el mercado nacional, y tal vez, en el futuro, llegar a ser una gran compañía distribuida a lo largo del territorio y comenzar el proceso de exportación solo con capital nacional.

**Palabras clave:** agave, consejo regulador del tequila, exportación, marcas, tequila.

## Abstract

The objective of this research was to determine the degree of knowledge that the inhabitants of the Guadalajara Metropolitan Area (made up of the municipalities Guadalajara, Tlajomulco de Zúñiga, Tlaquepaque, Zapopan and Tonalá) had regarding tequila and the brands produced in Los Altos of Jalisco. For this, a survey consisting of five questions was designed, which was applied in the central square, center or zócalo of each municipality. The results show that the big brands, when acquired by international companies, focused their attention on capturing the consumer in international markets, since the prices of the same products that they export have been out of the pocket of those who like That drink, so it could be considered that the big brands, have left the national market a little behind, of course they did not abandon it completely, but it stopped being their main objective. Therefore, it can be concluded that the national market is the window of opportunity to join the small and still unknown producers to work together and in a grouped way, they are able to standardize a series of products that being of the same quality and same packaging, they can cover the national market, and perhaps, in the future, become a large company distributed throughout the territory and begin the export process only with national capital.

**Resumo**

O objetivo desta pesquisa foi determinar o grau de conhecimento que os habitantes da Região Metropolitana de Guadalajara (composta pelos municípios Guadalajara, Tlajomulco de Zúñiga, Tlaquepaque, Zapopan e Tonalá) tinham sobre a tequila e as marcas produzidas em Los Altos. Jalisco. Para isso, foi elaborado um questionário composto por cinco questões, que foi aplicado na praça central, centro ou zócalo de cada município. Os resultados mostram que as grandes marcas, quando adquiridas por empresas internacionais, focaram sua atenção na captação do consumidor nos mercados internacionais, já que os preços dos mesmos produtos que exportam têm saído do bolso de quem gosta dessa bebida, então pode-se considerar que as grandes marcas, deixaram o mercado nacional um pouco para trás, claro que não o abandonaram por completo, mas deixou de ser o seu principal objetivo. Portanto, pode-se concluir que o mercado nacional é a janela de oportunidade para unir os pequenos e ainda desconhecidos produtores para trabalharem juntos e de forma agrupada, sendo capazes de padronizar uma série de produtos que sejam da mesma qualidade e mesmas embalagens. , podem cobrir o mercado nacional, e talvez, no futuro, se tornar uma grande empresa distribuída por todo o território e iniciar o processo de exportação apenas com capital nacional.

**Palavras-chave:** agave, conselho regulador da tequila, exportação, marcas, tequila.

# Introducción

Casi todos los mexicanos han escuchado la palabra *tequila* por lo menos una vez en su vida. Algunos la relacionan únicamente con la bebida alcohólica, otros con la población que se ubica en la región de los Altos de Jalisco y que lleva dicho nombre; sin embargo, resulta increíble que alrededor del mundo cada vez más personas conocen dicha palabra y la asocian a una bebida alcohólica de gran calidad. En el extranjero podrán ignorar que su nombre lo adoptó de una población, pero lo identifican con calidad y México.

De hecho, esta bebida ha llegado a ser tan consumida que algunos inversionistas han impulsado el crecimiento de las tequileras, lo que ha beneficiado a las poblaciones del estado de Jalisco, las cuales reciben los beneficios de los empleos directos e indirectos que genera este producto. Además, debido al incremento excepcional en cuanto a exportaciones que ha tenido en los últimos 10 años, ahora resulta habitual que en muchos países se ofrezca el tequila en reuniones o se adquiera directamente en tiendas importadoras de licores, así como en restaurantes y bares. Esto se debe a las estrategias de mercadeo usadas para dar a conocer la calidad de las diferentes presentaciones del tequila, lo que se ha constituido en un aporte significativo para el producto interno bruto de México.

La expansión y crecimiento que ha tenido la aceptación del tequila en el mundo es una ventana de oportunidades para que los pequeños productores (que no forman parte de los grandes consorcios) puedan establecer un sistema colaborativo de producción y distribución de sus propios productos para el mercado interno.

## Marco teórico

El tequila es una bebida tradicional mexicana que junto al mariachi y a la charrería se convierten por antonomasia en el símbolo representativo de México en el mundo. Su origen combina técnicas de preparación europea con plantas endémicas americanas. A través de los años, el número de tequileras ha creciendo de manera significativa hasta rebasar las 1400, aunque vale acotar que deben estar localizadas solo en alguno de los 181 municipios donde se cuenta con la debida acreditación para elaborarlo y para darle dicho nombre, es decir, Jalisco (125), Michoacán (30), Tamaulipas (11), Nayarit (8) y Guanajuato (7) (Cadeño, 2018).

Desafortunadamente, solo unas cuantas tequileras tienen la capacidad de exportarlo exitosamente (Carrillo, Pérez y Romero, 2010). Para obtener esa calidad, las tequileras se encuentran organizadas por el Consejo Regulador del Tequila (CRT), organismo que se dedica a verificar y certificar el cumplimiento de la Norma Oficial Mexicana (NOM) para el tequila, así como a promover la calidad, la cultura y el prestigio de la bebida nacional, por lo que le fue asignada la NOM-006-SCFI-2012 para establecer sus condiciones de fabricación. De hecho, al ser cumplido el anterior requisito se le considera como producto con garantía de calidad, el cual puede ser definido del siguiente modo de acuerdo con el CRT:

> "Bebida alcohólica regional obtenida por destilación de mostos, preparados directa y originalmente del material extraído, en las instalaciones de la fábrica de un Productor Autorizado la cual debe estar ubicada en el territorio comprendido en la Declaración, derivados de las cabezas de Agave tequilana weber variedad azul, previa o posteriormente hidrolizadas o cocidas, y sometidos a fermentación alcohólica con levaduras, cultivadas o no, siendo susceptibles los mostos de ser enriquecidos y mezclados conjuntamente en la formulación con otros azúcares hasta en una proporción no mayor de 49% de azúcares reductores totales expresados en unidades de masa, en los términos establecidos por esta NOM y en la inteligencia que no están permitidas las mezclas en frío. El Tequila es un líquido que, de acuerdo a su clase, es incoloro o coloreado cuando es madurado o cuando es abocado sin madurarlo" (NOM-006-SCFI-2012).

Su historia inicia hace cuatro siglos y toma su nombre de la población de Santiago de Tequila, ubicada en la región de los Altos de Jalisco, caracterizada por su tierra rojiza (Munguía, 1984). A través del tiempo, el tequila ha tenido distintas oportunidades frente a las bebidas alcohólicas elaboradas en otros países, y con el tiempo se convirtió en una industria estratégica de alto impacto en el desarrollo económico de los Altos de Jalisco, en donde da empleo a casi 70 000 personas y aporta 4200 millones de pesos anuales de recursos fiscales a través del Impuesto Especial sobre Producción y Servicios (IEPS); además, genera divisas por 1300 millones de dólares a través de sus exportaciones (Romo, 2018).

## Desarrollo

La denominación de origen es un instrumento, enmarcado en el derecho de propiedad industrial, que protege productos alimenticios y bebidas mediante la referencia de su lugar de origen que le confiere en el mercado características que suponen una garantía de calidad específica (Silverio, 2018).

El tequila es el resultado de destilar (técnica europea) el mosto fermentado obtenido del corazón de la planta (americana) conocida como *agave azul*, semejante a una piña gigantesca, que se conoce con el nombre *mezcal*, que en náhuatl significa 'la casa de la luna'. Existen 200 tipos diferentes de agaves, por lo que en diferentes lugares se obtienen bebidas aguardentosas similares al tequila, las cuales reciben el nombre de *mezcal*. Estas toman el apellido de la población donde nacen, de ahí que exista el mezcal de Oaxaca, de Cotija, de Quitupán, de Tonaya, de Tuxcacuesco, de Apulco, etc., aunque el más famoso es el de Tequila, el cual nace durante la época colonial en la población de Santiago de Tequila, fundada el 15 de abril 1530. Allí, en 1600, D. Pedro Sánchez de Tagle, marqués de Altamira, funda la primera fábrica productora de tequila, la cual introdujo el cultivo y destilación del mezcal, lo que le dio nombre a la población que está localizada a 58 kilómetros de Guadalajara con rumbo al puerto de San Blas en Nayarit, en la costa del Pacífico. Este territorio perteneció al Corregimiento de Tequila de la Audiencia de Nueva Galicia. En él se da muy bien el agave azul. De hecho, hoy en día se yerguen pequeñas y grandes fábricas del famoso licor, el cual antes de la simplificación publicitaria era conocido como *vino de mezcal de Tequila* (Jiménez, 2009).

No hay que pasar por alto el hecho de que la planta de maguey siempre ha sido base para producir el tequila, la cual es de gran importancia para la vida cotidiana de sus pobladores, pues sus hojas se aprovechan para construir techumbres, fabricar agujas, punzones, alfileres y clavos, hacer cuerdas, elaborar papel y ciertos recipientes. Además, sus pencas secas se usan como combustible, sus cenizas se empleaban como jabón, lejía o detergente, y su savia servía para curar heridas.

## La participación histórica del tequila

Al consumarse la Independencia en 1821, los vinos y licores españoles tuvieron dificultades para llegar a México, lo que dio oportunidad a las tequileras para incrementar sus ventas en Guadalajara, en el centro del país y en la Ciudad de México. Buscando expandir sus mercados, se aprovechó su cercanía al puerto de San Blas (204 km por la carretera actual) y la fiebre del oro en California en 1849, ya que la distancia entre San Blas, Nayarit y San Francisco, C. A. (2478 km por mar), comparada con la distancia de Nueva York a San Francisco, C. A., sería menos de la mitad (actualmente por tierra son 4700 km), y tomando en consideración que fue el territorio anexado a Estados Unidos en 1848 y se encontraba poblado por una gran cantidad de gente de ascendencia mexicana.

En 1857, estalló la guerra de Reforma para terminar el orden social heredado de la dominación española; por su parte, los productores de tequila apoyaron a los liberales, pensando en el futuro de su industria. A fines del siglo XIX e inicios del XX, el tequila tuvo como enemigo al ferrocarril norteamericano que llevaba con facilidad los aguardientes europeos de costa a costa.

Sin embargo, entre las clases bajas estaban los bebedores de aguardiente, por lo que el tequila se fue incrementando paulatinamente hasta lograr ventas considerables. Después de terminar con la dictadura y guerras revolucionarias, el país entero se volcó a buscar expresiones y costumbres para fortalecer la nacionalidad mexicana. Beber tequila en vez de aguardientes importados fue una gesta, por lo que el gobierno favoreció la conciencia de imagen del tequila como un símbolo nacional.

Luego, la industria cinematográfica mexicana de las décadas de 1930 y 1940 también contribuyó a este fin creando el falso estereotipo del macho mexicano, lo que influyó a acrecentar la fama de la bebida. Para atender la creciente demanda, se dispuso de pequeñas botellas fabricadas en la industriosa ciudad de Monterrey, con lo que se evitó su distribución a granel y en grandes barricas.

Durante la Segunda Guerra Mundial, dejó de llegar el whisky de Europa a Estados Unidos, por lo que el tequila alcanzó niveles insospechados de exportación al suplirlo. No obstante, al término del conflicto bélico —lo que originó la caída abrupta de su exportación— hubo que hacer un gran esfuerzo para incrementar el consumo nacional y para buscar simultáneamente los mercados de Europa y Sudamérica.

Para 1950, la producción de tequila gozó de mejoras técnicas considerables. Muchas fábricas, sin detrimento de su calidad, alcanzaron altos índices de rendimiento e higiene; además, algunas marcas fueron accesibles para gargantas comunes por ser de menos graduación, aunque en varios países se falsifica el tequila sin que sus gobiernos se preocupen por ello, a pesar de existir convenios y acuerdos internacionales (destaca el de Lisboa).

Los campos agaveros están en la franja central de Jalisco y su relevancia es muy importante debido a que la industria emplea alrededor de 300 000 personas comprometidas y orgullosas de participar directa o indirectamente en la fabricación del producto imbricado profundamente en la vida de la región occidental de México (CRT, 2019).

## Comprobar que se trata de tequila

Es importante saber que existen varias clases de tequila, el *blanco* es transparente o casi transparente; el *joven* u *oro* es de color ámbar o dorado; el *reposado* es de color pálido; el *añejo* es ámbar oscuro, y el *reserva* es de color más intenso. La forma tradicional de tomarlo es usando un caballito, y para catarlo se usa una copa.

Independientemente del color blanco o añejo, se puede observar cómo se forma un halo (corona) en la orilla de la copa. Para ver su cuerpo, se agita e inclina un poco la copa para que a las paredes se adhiera el tequila. Se forman lágrimas (gotas) que se deslizan por la copa. Si es ligero, medio o entero depende de la rapidez de las gotas al descender. Su aroma se percibe al colocar la nariz en el centro de la copa. Antes del sabor, se debe enjuagar la boca con un trago de agua. Se da un pequeño sorbo, se pasa a la boca, se retiene por diez segundos y se traga; después el aire se exhala por la nariz (Rodríguez, 2 de marzo de 2017).

## Investigación en la tequilera

La intervención de manos mexicanas en el proceso de elaboración es de vital importancia, ya que son tradiciones transmitidas de padres a hijos por generaciones; además de su calidez, la tierra en donde crece el agave le da ese sabor preferido por millones de exigentes paladares en el mundo. Por ello, se eligió visitar la tequilera Don Julio, en Atotonilco el Alto, por el importante papel de su tequila en el mundo.

Comparada con otras tequileras, el linaje de Don Julio tiene casi 80 años de haber visto la luz en 1942 con don Julio González, quien hizo su primer tequila a los 17 años. Luego, en 1947, abre su destilería en Atotonilco, Jalisco. Inicialmente, solo tenía como objetivo producir tequila en pequeñas cantidades para su reserva personal. Pero su calidad, artesanía y sabor se expandieron hasta llegar a todo México para convertirse en el primer tequila de lujo que goza del mayor reconocimiento nacional. El tequila tiene un proceso de producción que es único. Inicia con el cultivo del agave azul weber y la fabricación de las barricas de roble blanco americano que se usan para añejarlo; su elaboración requiere de habilidosas y experimentadas manos de artesanos.

Tequila Don Julio está elaborado con 100 % de agave azul weber (Intagri, 2018) que se debe cosechar manualmente. La planta de agave toma un periodo de tiempo que va siete a diez años para ser jimada, lo que le da su característico y suave sabor maduro del agave. Don Julio se distingue por su calidad y elaboración artesanal, de ahí que sea considerado el tequila más fino del mundo.

Gran parte del éxito mundial de esta tequilera se debe a que fue adquirida en 2019 por la empresa británica Diageo (Celis, 25 de marzo de 2019), que también lo distribuye a 60 países y tiene como objetivo alcanzar a otros, así como reforzar su presencia en Estados Unidos.

Luego de que Diageo adquiriera la tequilera Don Julio, en México la marca creció casi tres veces más rápido, y aumentó 13 % sus ventas en el país, pues su objetivo era ganar a la Casa Cuervo, líder en ventas a nivel nacional e internacional (*Latitud 21*, 31 de julio de 2017). La producción de tequila Don Julio supera los 10 millones de litros anuales y el 55 % se exporta. Por su parte, Diageo ofrece sus productos en más de 180 países, por lo que se prevé que aumentará la producción y exportación; asimismo, se procurará incrementar las ventas de 300 000 a 500 000 cajas en México y alcanzar el millón en exportaciones. Por ello, está invirtiendo más de 400 millones de dólares en cinco años en la construcción de una nueva destiladora y ampliación de la actual. Por lo anterior, Don Julio es la tequilera más importante de la región de los Altos de Jalisco y compite con la casa José Cuervo por ser la mejor del país (Rodríguez, 23 de mayo de 2017).

## Las exportaciones

La tabla 1 muestra la cantidad de litros de alcohol exportado a nivel mundial (Estados Unidos fue el mayor importador de este producto en 2019) (Mendoza, 27 de febrero de 2020).

**Tabla 1.** Los 10 países que importaron más tequila en 2019

| País consumidor | Millones de litros de alcohol |
|---|---|
| Estados Unidos | 204 443.07 |
| Alemania | 5052.20 |
| España | 3725.31 |
| Francia | 3559.90 |
| Japón | 2290.72 |
| Canadá | 2036.48 |
| Reino Unido | 2028.66 |
| Letonia | 1729.35 |
| Italia | 1660.88 |
| Colombia | 1370.43 |
| Total | 227 896.98 |

Fuente: Elaboración propia con los datos del Concejo Regulador del Tequila (2019)

**Figura 1**. Los tequilas más pedidos a nivel mundial en 2017

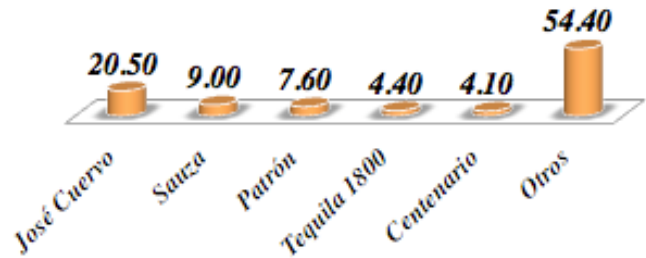

Fuente: Elaboración propia con datos del Concejo Regulador del Tequila

De acuerdo con el CRT, los datos de las exportaciones de tequila indican que durante el año 2017, de acuerdo con su reporte presentado en 2017, la Casa José Cuervo (figura 1) tuvo las mayores ventas que la colocan como la más grande del mundo y así permaneció ya que en el 2018, sus exportaciones se incrementaron en 3,8% con lo que llegó a 21,3 millones de cajas de 9 nueve litros y ventas netas de 28 158 millones de pesos.

**Figura 2.** Consumo de bebidas alcohólicas por marcas en el mundo

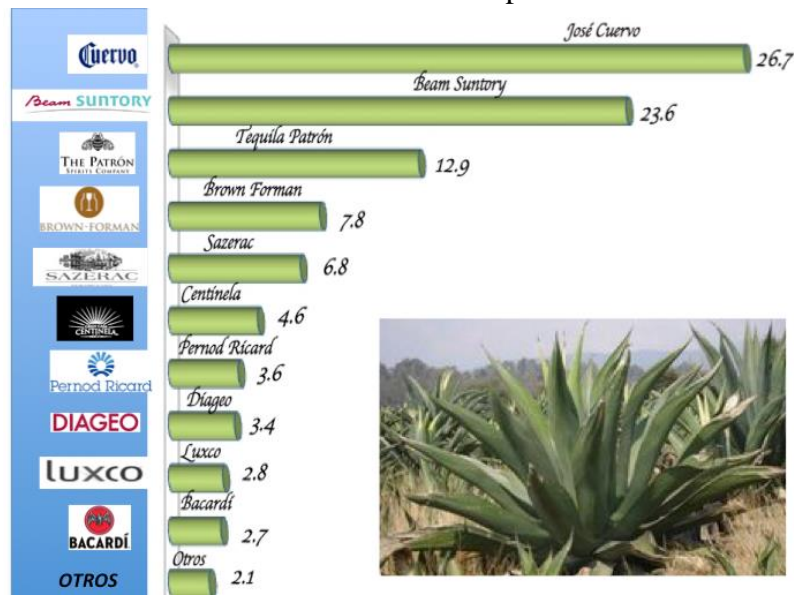

Fuente: Elaboración Propia con datos de *Orús (2018)*

La figura 2 muestra algunas de las bebidas alcohólicas mas reconocidas que se distribuyen a nivel mundial: la Casa José Cuervo, Beam Suntory y Tequila Patrón satisfacen 50.4 % del mercado.

Por otra parte, Euromonitor (figura 3) señala que las empresas con más ventas en todo el mundo de cajas de tequila de nueve litros en 2018 fueron las siguientes: Sauza (9.7 millones), seguida de Patrón (2.5 millones), Don Julio (1.5 millones), El Jimador (1.3 millones), Hornitos (1.3 millones) y Olmecas Altos (1.1 millones) (Ramírez, 15 de enero de 2020).

**Figura 3.** Exportación del tequila (millones de cajas de 9 litros) en 2018

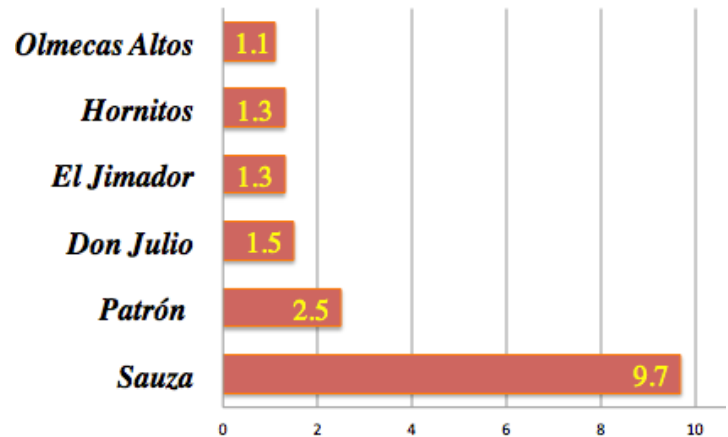

Fuente: Propia a partir de los datos Ramírez en Euromonitor (2020)

Debido a la tendencia de crecimiento de la industria tequilera, nueve grandes marcas, de las 148 reconocidas por el CRT, fueron compradas por empresas trasnacionales, mientras que el resto es de capital 100 % mexicano. En tal sentido, las trasnacionales que adquirieron las tequileras son las siguientes: la inglesa Allied Domecq, propiedad de la estadounidense Beam Future Brands compró el 100 % de Tequila Sauza; Pernord Ricard de Francia adquirió Tequila Viuda de Romero y comercializa la marca Olmeca; Bacardí de Bermuda es dueña de Tequila Cazadores; Brown Forman Corp. de EE. UU. posee Tequila Herradura, y Diageo de Inglaterra es dueña de Don Julio y de más de la mitad de Casa Cuervo. Entre ambas tiene un gran porcentaje de las exportaciones de tequila, lo que genera empleos e impuestos por las exportaciones. Con la adquisición de las grandes tequileras, México dejó de ser el principal mercado para el tequila, ya que 7 de cada 10 litros producidos se consumen fuera del país (Sánchez Fermín, 2019)

Del total de exportaciones, Estados Unidos recibe más de 80 % del tequila producido, lo que lo convierte en el principal consumidor mundial, ya que en 2016 en ese país se consumieron 161 millones de litros, es decir, 94 % más que lo consumido en 2005 (Sánchez Fermín, 2019).

Por su volumen de exportación, el tequila se ha convertido en una industria mexicana generadora de divisas por los impuestos de ventas**.** En 2016, las exportaciones de tequila de Jalisco fueron de 1.2 billones de dólares, 1.4 % más que en 2015. Asimismo, el número de empleos directos en esta industria llegó a 70 000 y el número de marcas nacionales registradas y certificadas que comercializan tequila ascendió a 1407 (Sánchez Fermín, 2019).

## Metodología

El objetivo de la presente investigación fue determinar el grado de conocimiento que los pobladores de la Zona Metropolitana de Guadalajara (integrada por los municipios Guadalajara, Tlajomulco de Zúñiga, Tlaquepaque, Zapopan y Tonalá) tenían en cuanto al tequila y a las marcas producidas en los Altos de Jalisco. Para ello, se diseñó una encuesta conformada por cinco preguntas, la cual fue aplicada en la plaza central, centro o zócalo de cada municipio. En uno de los municipios que componen la Zona Metropolitana de Guadalajara, se seleccionaron de forma aleatoria a 70 personas mayores de edad que manifestaron su gusto por las bebidas alcohólicas, los resultados que se obtuvieron de la aplicación de esas encuestas, fueron los datos que se procesaron para determinar cuánta gente (de la encuestada) conocía las marcas que se le presentaron.

Ya que de las respuestas que se obtuvieran, se estaría planteando el grado de oportunidad que se tendría al reunir a los pequeños productores en una sola marca para satisfacer el mercado nacional.

La primera pregunta fue la siguiente: *De las siguientes tequileras, ¿cuál es la empresa de la que has escuchado más?* Se propusieron las cuatro empresas más importantes de la Región de los Altos, es decir, Don Julio, 7 Leguas, San Matías y Patrón. De los cinco municipios que componen la Zona Metropolitana de Guadalajara (ZMG), se obtuvieron un total de 350 personas que apoyaron contestando la encuesta, y los resultados que se obtuvieron se muestran en la figura 4. Véase que la empresa Tequilera Don Julio, ubicada en Atotonilco el Alto, Jalisco, es parte importante del patrimonio de la Región de los Altos y la más conocida por sus productos, ya que el 79 % de los encuestados así lo respondieron.

Con los resultados que se obtuvieron, se muestra claramente que en la gente aficionada al consumo del tequila en la ZMG, mayoritariamente con el 79% sabe que es Don Julio la marca que es mas conocida en la región.

**Figura 4.** ¿Cuál es la empresa de la que has escuchado más?

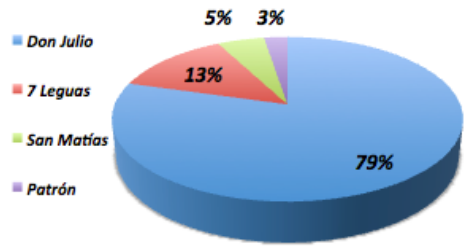

Fuente: Elaboración propia

La segunda pregunta de la encuesta procuró descubrir la cantidad de personas que saben de la calidad del tequila que se produce en la Región de los Altos de Jalisco. Al respecto, vale acotar que cuando se desconoce la calidad del tequila, solo basta ver la etiqueta del producto, pues en ella se debe encontrar la denominación de origen, que certifica que es tequila elaborado de agave azul; por ello, debe provenir de lugares específicos de México como Jalisco, Tamaulipas, Michoacán, Guanajuato y Nayarit. Asimismo, se debe indicar el porcentaje de agave azul que contiene, que puede variar del 51 % a 100 % (figura 5).

**Figura 5.** Porcentaje de personas que contestó que en la Región de los Altos de Jalisco se produce tequila de muy buena calidad

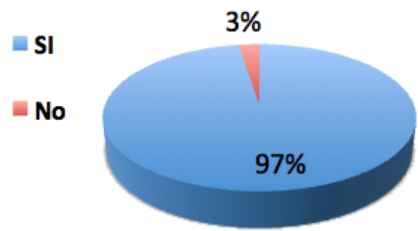

Fuente: Elaboración propia

Los resultados obtenidos muestran que tal vez por la tradición mexicana del cine nacional de antaño cuando se impulsó el tequila, lo mostró como un producto de la Región de los Altos de Jalisco y desde entonces, todo el tequila lo relacionan con esa zona del estado, sin pasar por alto el hecho de que recibe el nombre de la población ubicada en la misma zona.

La calidad del tequila depende de la casa productora y del tipo de añejamiento, que pueden ser *blanco* de color transparente (joven y sin añejar), *joven* y *oro* (mezcla de blanco con reposado o añejado), *reposado* (añejado como mínimo dos meses), *añejo* (oscuro y añejado al menos un año), *extra añejo* (sabor a madera, miel o vainilla, añejado tres años en promedio) y *reserva* (color y sabor más intenso, añejado por más de ocho años) (Díez, 3 de abril de 2017). En tal sentido, la tercera pregunta de la encuesta estuvo orientada a conocer si los precios del tequila corresponden a la calidad del producto adquirido, es decir, si la relación entre calidad y precio es adecuada. La figura 6 muestra que 80 % de los encuestados considera que en el tequila existe la relación calidad-precio, lo que lo pone al alcance del consumidor; por ello, se puede encontrar un tequila de buena calidad a un buen precio en el mercado.

**Figura 6**. Porcentaje de personas encuestadas que cree que el costo del tequila corresponde a su calidad.

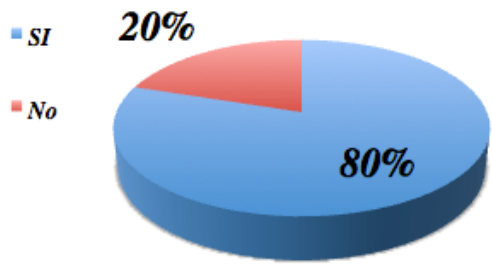

Fuente: Elaboración propia

Con estos datos sobre el costo y calidad del tequila, se puede apreciar que la gente de forma general (el 80% de los encuestados así lo manifestó), sabe que para paladear un tequila de muy buena calidad, tendrá que pagar un costo alto y hasta muy elevado en algunas ocasiones, motivo por el cual sabe que si cuesta poco o muy poco, no espera encontrar algo muy bueno y a veces ni siquiera bueno.

Cuando se adquiere una botella de tequila, se toman en consideración cuatro factores del producto: marca, precio, calidad y presentación. Sin embargo, como se observa en la figura 7, solo tres de estos factores son los que los consumidores toman en cuenta:

**Figura 7.** Factores que determinan la compra del tequila

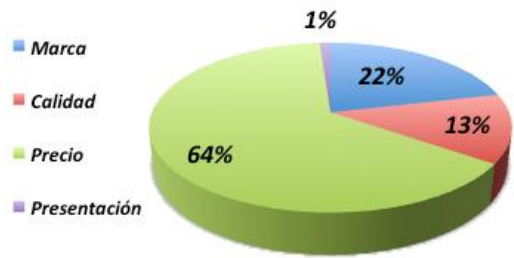

Fuente: Elaboración propia

La gente encuestada mostró con sus respuestas que la gran mayoría de ellos (el 64%), al adquirir un producto, lo primero en que se fija es el precio que tiene, y luego, si lo accesible, ve la marca y a continuación la calidad que viene asentada en la etiqueta y casi nadie toma como referencia para la adquisición, la presentación del producto.

La quinta y última pregunta de la encuesta tenía la intención de precisar si los encuestados consideran que debe de incrementarse la exportación a más países para llegar a una mayor cantidad de paladares de distintas nacionalidades. En esta pregunta contestaron 343 personas afirmativamente y solo siete lo hicieron de forma negativa (figura 8), lo cual indica en que la gran mayoría de las personas (98% de los encuestados) está de acuerdo en que esta bebida tradicional llegue a mas países y que más gente conozca un producto de alta calidad como lo es el tequila.

**Figura 8.** ¿Es importante incrementar la exportación del tequila?

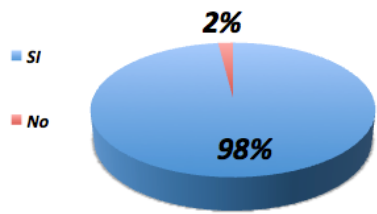

Fuente: Elaboración propia

Como se ha señalado en las páginas anteriores, el tequila ha tenido un gran auge en otros países debido a distintos factores relacionados, por ejemplo, con la calidad del producto, las compañías que han invertido en la adquisición de las tequileras para hacerlas crecer en producción, el personal que interviene en todo el proceso de producción y exportación, las estrategias de *marketing* implementadas, etc. Sin embargo, vale destacar que ese auge internacional no ha sido proporcional en el mercado interno.

Por ello, se ha venido trabajando (como parte de las instituciones educativas de las regiones en que se encuentran los nuevos campos agaveros del estado de Jalisco) con los pequeños productores de tequila, ya que reúnen las condiciones para que sus productos sean catalogados con la denominación de origen. Para eso, se han buscado estudiantes que provengan de familias de agaveros que conozcan de cerca los procesos y la mayoría de las actividades que están relacionadas con el producto.

El resultado de esa búsqueda es bastante alentador, pues en los centros educativos hay un número significativo de estudiantes tienen cualidades que pueden ser aprovechadas para la producción del tequila; es decir, están relacionados con procesos de compra-venta, cursan carreras de nivel medio-superior y superior asociadas con el desarrollo de sistemas computacionales y *marketing*, están familiarizados con las actividades del campo, etc. De hecho, algunos estudiantes provienen de familias económicamente solventes, por lo que cuentan con los recursos para fungir como inversores en nuevas empresas. Además, algunas personas adquirieron pequeñas extensiones de tierra (de una a cinco hectáreas) en las laderas de algunas depresiones de las sierras (p. ej., la sierra del Tigre o la sierra del Halo), donde han creado sus propias tequileras, aunque sus productos generalmente se usan para el consumo personal o para su venta en las localidades de la región.

Asimismo, en cuanto al personal de las instituciones educativas, se puede decir que cuentan con la preparación académica para conjuntar los elementos que participan en la estandarización de metodologías para que los inversores, productores de agave, estudiantes y profesores puedan asesorar a esas personas para establecer empresas que sean capaces de elaborar el tequila con distintas calidades. Incluso, se puede usar el conocimiento disponible en cuanto a áreas de *marketing*, redes sociales, facturación, seguimiento del producto, etc., lo cual serviría para poner en práctica las destrezas adquiridas e impulsar el crecimiento de la región de donde provienen.

El objetivo es conseguir que las pequeñas empresas fabriquen un número determinado de tequilas de distintas calidades y con precios más bajos que permitan competir con las grandes marcas para abastecer al mercado nacional. Esto, lógicamente, con base en las Normas Oficiales del Tequila (17025, 17020, 17065 y 14065) y con el apoyo del Consejo Regulador del Tequila.

## Conclusiones

Como conclusión se puede afirmar que un producto como el tequila, de origen humilde y surgido de la combinación entre una planta mexicana y un proceso español, ha crecido de manera tan impresionante que ha ocupado un lugar preferencial a nivel mundial. Por ello, estar fuera de México ya no es impedimento para degustar el sabor del tequila, ya que se puede encontrar en licorerías, tiendas de conveniencia, y bares y restaurantes de muchos países. Esto es posible porque la región de los Altos de Jalisco se ha visto muy favorecida con la multiplicación de las tequileras, ya que de acuerdo con los datos de la encuesta aplicada, las maracas que se incluyeron en dicha encuesta, el las respuestas que se obtuvieron dicen que un gran porcentaje de la población de la Zona Metropolitana de Guadalajara las conoce por lo que la gran mayoría señaló a Don Julio como la mas conocida, así también, esa gran mayoría estuvo de acuerdo que son tequilas de gran calidad los producidos en la región de los altos, al igual que relacionan esa calidad con los costos y que la inmensa m lo que ha favorecido a la exportación y, en consecuencia, al aumento de empleos directos e indirectos en la zona, así como a la economía del país.

Es cierto que a lo largo del tiempo ha tenido altas y bajas, como cualquier empresa en el mundo, pero siempre ha habido emprendedores que no han cejado en su empeño de sobresalir entre los demás. En otras palabras, los tequileros nunca se han dado por vencidos en las épocas desfavorables, y en cambio buscaron la forma de permanecer en el gusto del público para seguir creciendo.

En tal sentido, se puede indicar que las casas productoras más grandes están localizadas en las siguientes poblaciones: Atotonilco, con las tequileras Patrón, Don Julio, Quiote, Artesanal de los Altos y 7 Leguas; Arandas con Centinela, Cazadores y El Charro; Tequila, con las tequileras José Cuervo, Orendain, Sauza y Viuda de Romero; Amatitán, con Herradura, y Magdalena con la tequilera San Matías. De ellas, algunas tienen la mayor exportación por haber sido adquiridas por compañías trasnacionales. Este aumento en las exportaciones ha sido motivante para que las empresas que adquirieron las plantas tequileras sigan invirtiendo para ampliar las instalaciones y modernizar procesos sin perder su calidad.

Por ello, en la actualidad se ha estado incursionando en la formación de un grupo tequilero colaborativo donde se reúnan, inicialmente, algunas de las tequileras pequeñas. El propósito es tratar de llegar a más consumidores del mercado nacional para, en el futuro, convertirse en empresas exportadoras con capital 100 % mexicano.

## Trabajo futuro

Muchas de las grandes empresas tequileras de la actualidad iniciaron produciendo la bebida para consumo personal o para sus allegados, pero gracias a la calidad del producto y al trabajo incesante de los dueños pudieron convertirse en los mayores exportadores del país.

Después de hacer una análisis de los resultados que se obtuvieron de aplicar las encuestas, se pudo apreciar que los encuestados, aprecian las calidades que les brinda el tequila como bebida que se toma en diferentes ocasiones, aunque también, permite detectar que debido a que el poder adquisitivo del mexicano en promedio, no le permite acceder a las mejores calidades del producto, de ahí que surgió la idea de producirlo para llegar a ese público.

Después de haber detectado las oportunidades que se tienen para producir el tequila con la misma calidad que el de exportación pero a precio que pueda ser adquirido por el mexicano promedio, se establece que el siguiente paso debe apuntar a la formación de una organización de profesores que brinde ayudas técnicas a los pequeños productores y que los relacionen con los dueños de los capitales que se encuentran en los municipios para que de esas alianzas surjan empresas capaces de trabajar de forma colaborativa con participación de los siguientes entes:

a) Capitalistas que puedan visualizar el potencial que tiene invertir en una industria que ayude a su población de origen o de residencia.

b) Agricultores que actualmente tengan pequeñas propiedades sembradas de maguey y que estén produciendo el tequila de forma artesanal, el cual se puede vender a un público reducido o para consumo personal.

c) Universidades que —a través de profesores y estudiantes— sean capaces de brindar su tiempo y conocimientos de distintas áreas para asesorar en todo el proceso de producción, es decir, adquisición de materiales y materias primas, tramitación de permisos, embalajes, transportación, comercialización, etc. Asimismo, involucrar a los estudiantes para que desarrollen actividades vinculadas con las carreras que están cursando y con este producto, pues de ese modo podrán poner en práctica la teoría aprendida en las aulas de clases.